# Bipolar conduction and giant positive magnetoresistance in doped metallic titanium oxide heterostructures

Ke Huang[1,+], Tao Wang[2,+], Mengjia Jin[3], Liang Wu[4], Junyao Floria Wang[5], Shengyao Li[1], Dong-chen Qi[6], Shuying Cheng[3], Yangyang Li[7], Jingsheng Chen[7], Xiaozhong He[2], Changjian Li[7], Stephen J. Pennycook[7], X. Renshaw Wang[1,8,*]

[1]*Division of Physics and Applied Physics, School of Physical and Mathematical Sciences, Nanyang Technological University, 21 Nanyang Link, 637371, Singapore*
[2]*Institute of Fluid Physics, CAEP, P. O. Box 919-106, Mianyang 621900, P.R. China*
[3]*College of Physics and Information Engineering, Institute of Micro-Nano Devices & Solar Cells, Fuzhou University, Fuzhou 350108, P. R. China.*
[4]*School of Material Science and Engineering, Kunming University of Science and Technology, Kunming, 650093, Yunnan, P. R. China*
[5]*Hwa Chong Institution (College), Singapore 269734*
[6]*ARC Centre of Excellence in Future Low-Energy Electronics Technologies, School of Chemistry, Physics, and Mechanical Engineering, Queensland University of Technology, Brisbane, Queensland 4001, Australia*
[7]*Department of Materials Science and Engineering, National University of Singapore, 117575, Singapore*
[8]*School of Electrical and Electronic Engineering, Nanyang Technological University, 50 Nanyang Ave, 639798, Singapore*

[+]These two authors contributed equally
*renshaw@ntu.edu.sg

**Keywords:** Bipolar conduction, Giant magnetoresistance, Deuterium doping, titanium oxide, nonlinear Hall effect

**Abstract**
Empowering conventional materials with unexpected magnetoelectric properties is appealing to the multi-functionalization of existing devices and the exploration of future electronics. Recently, owing to its unique effect in modulating a matter's properties, ultra-small dopants, *e.g.* H, D, and Li, attract enormous attention in creating emergent functionalities, such as superconductivity, and metal-insulator transition. Here, we report an observation of bipolar conduction accompanied by a giant positive magnetoresistance in D-doped metallic Ti oxide ($TiO_xD_y$) films. To overcome the challenges in intercalating the D into a crystalline oxide, a series of $TiO_xD_y$ were formed by sequentially doping Ti with D and surface/interface oxidation. Intriguingly, while the electron mobility of the $TiO_xD_y$ increases by an order of magnitude larger after doping, the emergent holes also exhibit high mobility. Moreover, the bipolar conduction induces a giant magnetoresistance up to 900% at 6 T, which is ~6 times higher than its conventional phase. Our study paves a way to empower conventional materials in existing electronics and induce novel electronic phases.

**Introduction**

Induced by coupling the charge with other degrees of freedom, *e.g.* spin, orbit, and lattice, emergent electronic phenomena spur new research interests and contribute to the future application[1–3]. For example, the unexpected two-dimensional electron gas[4] and gate-tunable superconductivity[5] arouse a mania in discovering the novel electrical properties, which seems unachievable in the original system. Among various fertile grounds for appealing electromagnetic behaviors, one tantalizing platform is the metallic system, where the charge may induce phenomena, such as charge-spin conversion[6] and charge density wave[7]. However, even though charge could be coupled, tailoring the charge carrier in a metallic system, including its type, density, and mobility, is still challenging. The major obstacle is that the high carrier density and mobility dominate the electrical transport and overwhelm any modulated electrical signals. Therefore, a successful modulation of carriers in a metallic system could reveal both the hidden physics and their implication in further electronics.

To achieve the tunability of carrier behavior in a metallic system, both the dopant and material system are crucial. Doping with ultra-small dopants, such as deuterium (D), is a possible solution based on its successful modulations in semiconductors[8,9]. Significantly, the applications of ultra-small dopants in the creation of metal-insulator transition[10,11] and high-temperature superconductivity[12,13] demonstrate the huge potential for future electronic applications by introducing strongly-correlated functionalities. Apart from a proper dopant, a metallic system with a subtle and tunable electronic structure is preferred. One promising material is titanium oxide ($TiO_x$), which has tunable oxidation states with the titanium ion charged from 2+ to 4+[14]. Moreover, $TiO_x$ is earth-abundant[15] and has been widely used in important applications, such as photovoltaics[16], photocatalysis[17], and ultraviolet absorption[18]. Particularly, in the rocksalt titanium monoxide ($Ti^{2+}O$), the $d_{xy}$, $d_{xz}$, and $d_{yx}$ orbits overlap and form the partially filled $t_{2g}$ orbit, leading to a unique metallic behavior [19,20], different from the other semiconducting $TiO_x$ families. Interestingly, rocksalt TiO could have a stoichiometric O/Ti ratio larger than 1[21], which could induce hole conduction[22].

In this work, we took advantage of both $TiO_x$ and ultra-small dopants to generate holes in the metallic $TiO_xD_y$. Structural, electronic, and stoichiometric analyses reveal that majority of the film maintains a rocksalt structure of Ti monoxide even after D doping. The electron mobility after a proper D doping significantly increases by an order of magnitude compared with the undoped one. Holes also exhibit comparably high mobility proved by fitting the nonlinear Hall effect with a two-channel model. Moreover, a giant positive magnetoresistance (MR) is observed and explained by the electron-hole compensation. Further investigation shows that the generation of holes and high carrier mobility can be tailored through controlling the D concentration.

**Results and discussion**

Figure 1a shows the schematic of the sample fabrication and resulting sample structure. Incorporating ultra-small dopants into oxides is challenging[9,23], but into metals is relatively well-investigated[24–26]. For the sake of accurate quantification, D is better than H, because distinguishing between H dopant and H contaminants from the environment is challenging while distinguishing D from H is relatively easy[27]. Therefore, we fabricate the $TiO_xD_y$ film with a two-step process, *i.e.* deuteration and oxidation. Experimentally, we sputtered 11 nm Ti films on $SrTiO_3$ (STO) substrates at 500 °C. The sputtering gas was a mixture of $D_2$ and Ar with four different $D_2$:Ar ratios, namely 0:1, 0.1:1, 0.3:1, and 0.5:1. Then, in the oxidation process, we utilized both the lattice oxygen from the STO[28,29] activated during the 500 °C growth and environmental oxygen from the atmosphere after the growth. The oxidation and deuteration conditions are detailed in Supplemental note 1 and 2 respectively. With sufficient temperature and long time oxidation, one expects a $TiO_xD_y$ thin film.

To evaluate the atomic and electronic structures of the $TiO_xD_y$, we conducted scanning transmission electron microscopy (STEM) and electron energy loss spectroscopy (EELS). Figure 1b shows a cross-section STEM image of the $TiO_xD_y$/STO with a sputtering gas ratio of $D_2$:Ar = 0.1:1. The majority of the film (~8 nm) shows a rocksalt structure with a lattice parameter ~4.18 Å, very close to that of the rocksalt TiO lattice parameter of 4.185 Å. This single crystallinity is unexpected because TiO and STO have a ~7% lattice mismatch. We hypothesize that the interfacial layer (~1 nm), the transition layer between the film and substrate, releases the lattice mismatch, and stabilizes the rocksalt structure of $TiO_xD_y$.

To investigate the Ti valence across the film, we carried out the EELS spectrum mapping at 17 different positions (marked in Figure. 1c) with an interval of 0.9 nm from the interface to the surface. Figure 1d,e shows the EELS results of Ti $L_{2,3}$ and O $K$ edge, respectively. In the STO (spectra #1- 4), Ti $L_{2,3}$ edges spectra show four peaks, a signature of $Ti^{4+}$. Correspondingly, a large O $K$ edge pre-peak at 528 eV is also indicative of oxygen bonded to $Ti^{4+}$ states. Moving into $TiO_xD_y$ film (from #5 onwards), Ti $L_{2,3}$ edge spectra has only two peaks and a significant redshift, implying that Ti is in a much lower valence state ($Ti^{2+}$/$Ti^{3+}$)[30–32]. The surface 2 nm region (#13-17) shows a slight blueshift, indicating a higher oxidation state. In comparison with the previous study[30] which offered a comprehensive list of $Ti_xO_y$ compounds, we confirm that mid-region of the film is predominantly $Ti^{2+}$, while the interfacial region (~1 nm) and surface region(~2 nm) are dominated by $Ti^{3+}$. Supplemental Figure 2 shows the D/Ti ratio in $TiO_xD_y$ film as a function of the sputtering gas ratio characterized by Rutherford backscattering (RBS) and elastic recoil detection analysis (ERDA). D is well introduced into the film even with the smallest $D_2$:Ar ratio = 0.1:1, and more D is doped when the $D_2$:Ar increases from 0.1:1 to as high as 10:1. Based on the structure and component characterization, we can conclude that the D doped TiO, with some deviation in the O/Ti atomic ratio, constitutes the majority of the film.

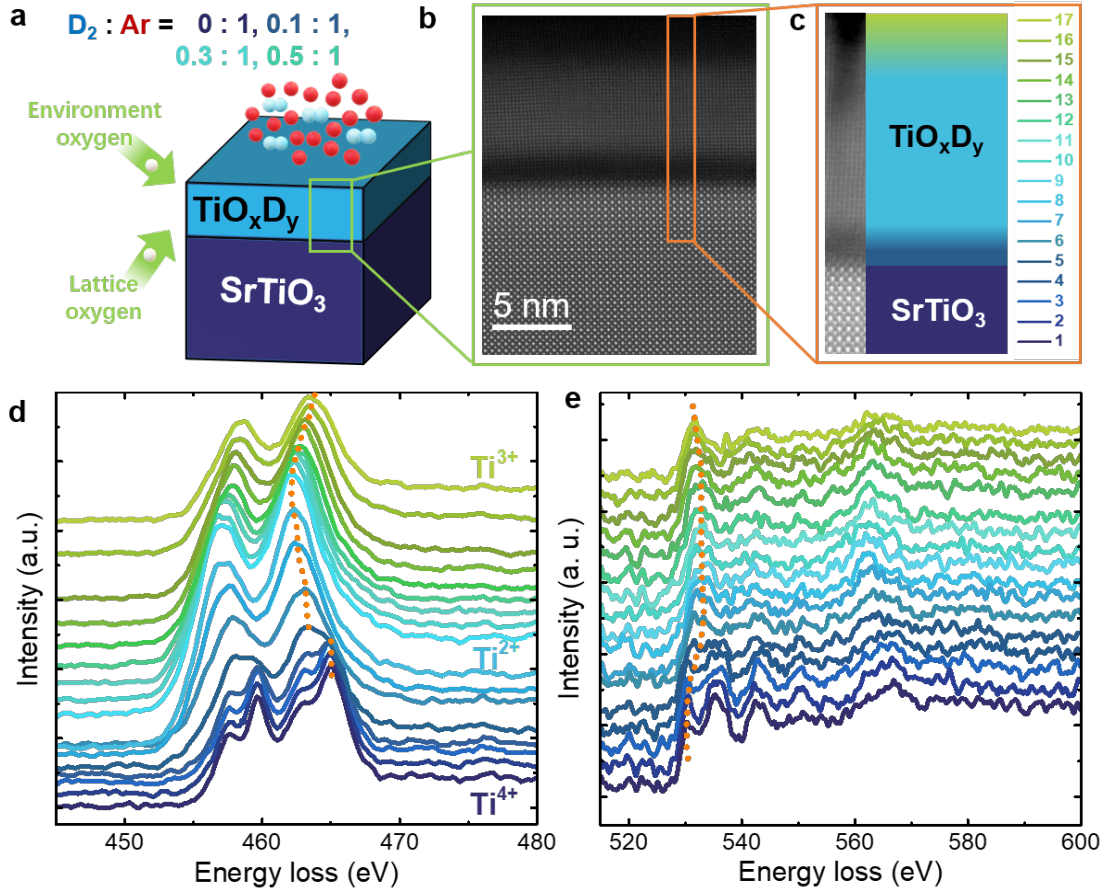

Figure 1. Scanning transmission electron microscopy (STEM) and electron energy loss spectroscopy (EELS) results of TiO$_x$D$_y$/STO sputtered in D$_2$:Ar = 0.1:1 (a) Schematic of TiO$_x$D$_y$/STO heterostructure, D doping, and oxidation. (b) High-resolution STEM image of the cross-section of TiO$_x$D$_y$/STO. (c) The high-angle annular dark-field (HAADF) image acquired simultaneously while taking the EELS spectrum imaging. The schematics image next to HAADF image denotes the relative positions of 17 EELS spectra of (d) Ti $L_{2,3}$ and (e) O $K$ edge.

Both previous work[20] and our temperature-dependent resistivity ($\rho$-$T$) (see Supplemental Figure 3) prove that TiO is metallic. To analysis on whether the D doping can tailor the carrier of the conductor, we measured the Hall resistivity ($\rho_{xy}$). In this work, both the $\rho_{xy}$ and $\rho$ were measured using a van der Pauw geometry. More details on the instruments used can be found in Supplemental note 3. Figure 2a-c shows the field-dependent $\rho_{xy}$ ($\rho_{xy}$-$B$) of TiO$_x$D$_y$ with D$_2$:Ar = 0.1:1 at different temperatures. $\rho_{xy}$ is linear field-dependent from 300 to 30 K, which corresponds to pure electron conduction. Strikingly, below 30 K, the $\rho_{xy}$-$B$ starts to exhibit a nonlinearity and reaches the maximum $\rho_{xy}$ at 10 K. From 10 to 0.3 K, the nonlinearity gradually diminishes. To analyze the nonlinearity, we fit the $\rho_{xy}$-$B$ with the two-channel model[33–35] shown in the following equation

$$\rho_{xy}(B) = \frac{Bn_1\mu_1^2 + B(n_2 + B^2(n_1+n_2)\mu_1^2)\mu_2^2}{e(n_1^2\mu_1^2 + 2n_1n_2\mu_1\mu_2 + (n_2^2 + B^2(n_1+n_2)^2\mu_1^2)\mu_2^2)} \quad (1),$$

where $e$ is the elementary charge; $B$ is the external field we applied; $n_1$ and $n_2$ are carrier density ($n$) of the first and the second conducting channel, respectively; $\mu_1$ and $\mu_2$ are the mobilities of the first and second conducting channel, respectively. In addition, during the fitting, $n_1$, $n_2$, $\mu_1$, and $\mu_2$ are constrained by the zero-field $\rho$ through the relation of $e(n_1\mu_1 + n_2\mu_2) = 1/\rho$. A detailed comparison between the fitting and the measurement results can be found in Supplemental Figure. 4,5. Figure 2d,e shows the fitted $n$ and mobilities, respectively. Above 30 K, the TiO$_x$D$_y$ film is pure electron-conducting. Interestingly, we observed a minimum $n$ value at ~100 K, which is very close to the reported temperature of STO's cubic-to-tetragonal phase transition, 105 K[36]. This indicates that the STO substrate contributes to the transport properties through the interface. At 30 K, holes are generated which corresponds to the nonlinear Hall effect. Below 30 K, the hole density ($n_h$) dramatically increases and exceeds electron density ($n_e$) at 3 K. Explaining the nonlinear Hall effect as a coexistence of electron and hole is consistent with previous studies[37,38]. Another exciting result is the high mobility of both carriers. Figure 2e shows that the fitted mobilities are up to around $4\times10^3$ cm$^2$V$^{-1}$s$^{-1}$ for holes and $2\times10^4$ cm$^2$V$^{-1}$s$^{-1}$ for electrons at 10 K. Note that the amorphous surface and interface of TiO$_x$D$_y$/STO dominated by Ti$^{3+}$ are insulating at low tempreture[20] and cannot hold such high-mobility carriers. The mid-region rocksalt TiO$_x$D$_y$ layer has an O/Ti ratio larger than 1 (see Supplemental note 1), demonstrating a cation vacancy. This vacancy arouses acceptor defects and can potentially exhibit hole conduction[22]. Moreover, the $n_e$ of the film at 300 K is close to the previous study on TiO[21]. Therefore, we conclude that the bipolar conduction in our film is the intrinsic properties of the rocksalt TiO$_x$D$_y$ layer.

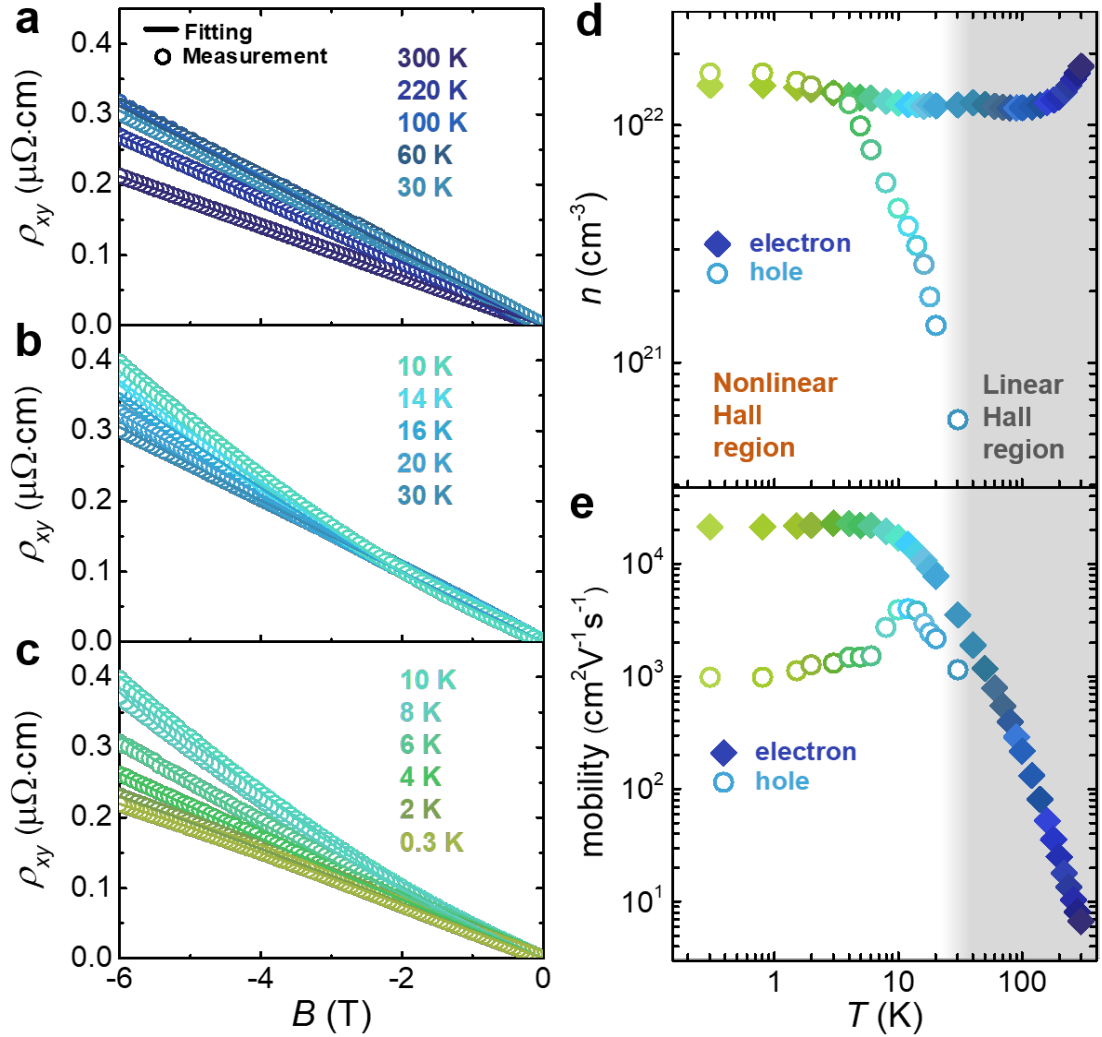

Figure 2. Nonlinear Hall effect of TiO$_x$D$_y$. $\rho_{xy}$ of TiO$_x$D$_y$ (D$_2$: Ar = 0.1:1) from 300 to 30 K (a), 30 to 10 K (b), 10 to 0.3 K (c). Circles and lines are the measurements and fittings, respectively. Fitted $n$ (d), and mobility (e) as a function of temperature. The coexistence of electron and hole below 30 K explains the nonlinearity of the $\rho_{xy}$ below 30 K.

To date, there is no report of the bipolar conduction with high-mobility carriers in titanium oxide yet[20,29]. Therefore, we need to confirm the relation between the hole and D dopants. We first compare the $\rho$-$T$ of TiO$_x$D$_y$ films with different D$_2$:Ar sputtering gas ratios. Figure 3a shows that all TiO$_x$D$_y$ films are metallic. This is consistent with the existing study on the metallic titanium monoxide[20]. At any temperature below 300 K, the $\rho$ decreases when we increase the D$_2$:Ar ratio. Quantitatively speaking, at both 300 and 1.5 K, the $\rho$ drops around an order of magnitude when the D$_2$:Ar ratio increases from 0:1 to 0.5:1. This huge $\rho$ drop confirms that D is doped into the film and proves that D has a significant impact on the film conductivity. To map this D enhanced conductivity onto bipolar conduction. We then compare the $\rho_{xy}$-$B$ (Figure 3b) and fitted the $n$ (Figure 3c top) and mobility (Figure 3c bottom). We first observed a giant electron mobility increase. TiO$_x$D$_y$ with D$_2$:Ar = 0.1:1 exhibits a high electron mobility, 2x10$^4$

cm$^2$V$^{-1}$s$^{-1}$ which is 10 times as large as the undoped one (D$_2$:Ar = 0:1). Moreover, TiO$_x$D$_y$ with D$_2$:Ar = 0.1:1 has the greatest degree of nonlinearity and the largest $n_h$ accordingly. Comparatively, the nonlinearity and the $n_h$ of the TiO$_x$D$_y$ with D$_2$:Ar = 0.3:1 is much smaller. The nonlinearity can be better demonstrated by comparing the dash lines (Figure 3b), the $\rho_{xy}$-B low-field extrapolations, with the $\rho_{xy}$-B curves. Despite the decreased $n_h$, hole mobility is relatively unchanged. When we further increase or decrease D$_2$: Ar to 0.5:1 or 0:1, the $\rho_{xy}$-B becomes completely linear. This indicates an optimal amount of D dopant is crucial for this nonlinearity and both excessive and insufficient D doping can inhibit the generation of holes.

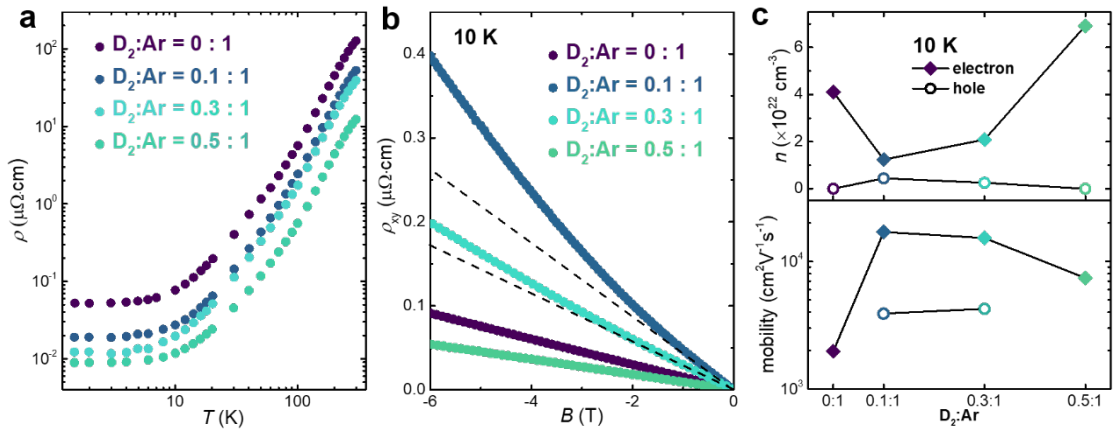

Figure 3. Control experiment to investigate the requirement for high-mobility bipolar conduction. $\rho$-T (a) and $\rho_{xy}$-B (b) of the TiO$_x$D$_y$ with different sputtering gas ratios (D$_2$:Ar). The dash lines are the low-field extrapolations, which are plotted to demonstrate the nonlinearity. (c) Fitted $n$ and mobility of TiO$_x$D$_y$ at 10 K. The one with D$_2$:Ar = 0.1:1 yields the highest hole density.

Following the carrier investigation, we measured the MR of TiO$_x$D$_y$ with D$_2$:Ar = 0.1:1 at different temperatures (see Supplemental Figure 6). We found an unexpected giant MR, 915% at 6 T, and 10 K. Such a giant positive MR is more than twice of that in either the heavily-doped STO[39] or the untreated TiO$_x$[40], which exhibit an MR less than 400% at 6 T. Typically, a giant unsaturated positive MR can be mainly be categorized into field-linear MR, observed in semiconductor[41,42] and field-nonlinear MR, observed in semimetal[43–45]. In the semimetal, the Fermi level crosses both the valance and conduction band, resulting in intrinsic bipolar conduction[44]. The electron and hole conduction channels then compensate each other and lead to an unsaturated large MR. Similar to the semimetal, the giant MR in our TiO$_x$D$_y$ samples may also relate to the bipolar conduction. Figure 4a compares the $\rho$-T at 0 and 6 T. Importantly, these two $R_s$-T curves coincide at the temperature region where the $\rho_{xy}$-B is linear (T > 30 K). The MR becomes significant only below 30 K, which is the exact transition temperature of $\rho_{xy}$-B from linear to nonlinear. Therefore, the giant positive MR is an intrinsic behavior of the bipolar conduction. This can be further proved by the comparison of the 10 K MR with different doping states (Figure 4b). When the D$_2$:Ar ratio is changed to 0.3:1, MR at 6 T is accordingly reduced to around 250%. The pure electron conduction $TiO_xD_y$ ($D_2$:Ar = 0:1 or 0.5:1) has an MR less than 150% at 6 T, which is 6 times smaller than the one of $TiO_xD_y$ with $D_2$:Ar = 0.1:1.

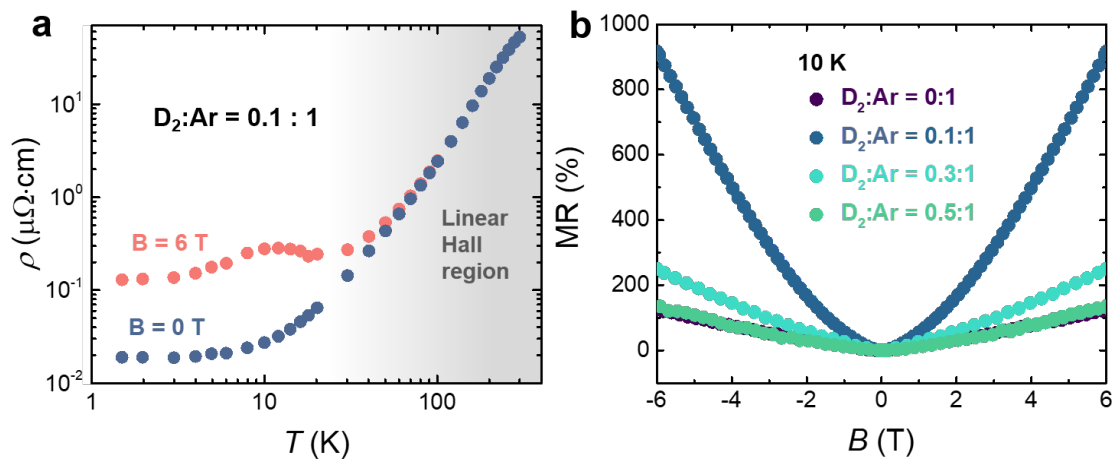

Figure 4. Giant positive magnetoresistance (MR) of $TiO_xD_y$. (a) $\rho$-$T$ with the out-of-plane external field set at 0 and 6 T. (b) $\rho_{xy}$-$B$ of the $TiO_xD_y$ with different sputtering gas ratios ($D_2$:Ar)

**Conclusion**

In this work, we successfully achieved unseen bipolar conduction in the metallic $TiO_x$ with an ultra-small dopant, D. The mobility of the electron is boosted 10 times after doping. The hole exhibits high mobility, and its density is comparable to the original electron. Both the hole generation and the high carrier mobilities can be tailored through controlling the doping conditions. This work validates the power of D doping in the carrier modulation. Significantly, through this modulation, realizing and tuning useful electronic phenomena, such as giant positive MR, are also possible.

**Acknowledgment**

X.R.W. design and directed the study. X.R.W. acknowledges supports from the Nanyang Assistant Professorship grant from Nanyang Technological University, Academic Research Fund Tier 1 (Grant No. RG177/18) and Tier 3 (Grant No. MOE2018-T3-1-002) from Singapore Ministry of Education, and the Singapore National Research Foundation (NRF) under the competitive Research Programs (CRP Grant No. NRF-CRP21-2018-0003). T.W. acknowledges supports from the National Natural Science Foundation of China under Grant No.11905206. C.L. and S.J.P. acknowledge the financial support from Singapore Ministry of Education AcRF Tier 2 fund MOE2019-T2-1-150.